\documentclass[aps,prd,showpacs,preprintnumbers,nofootinbib,twocolumn]{revtex4}
 
\usepackage{bm}
\usepackage{latexsym}
\usepackage{dcolumn}
\usepackage{amsfonts,amssymb}
\usepackage{graphicx,epsfig}
\usepackage{psfrag}
 
\def\beq{\begin{equation}}
\def\eeq{\end{equation}}    
\def\eeq{\end{equation}}
\def\br{\begin{eqnarray}}
\def\er{\end{eqnarray}}
\def\pa{{\partial}}
\def\l{\left}
\def\r{\right}
\def\pl{Painlev\'{e}}
\def\lem{Lema$\hat{\rm i}$tre}

\def\e{\epsilon}
\def\d{\partial}
\def\ast{{\em \/ Astron.\ J.~}}

\def\apj{{\em  \/Astro.\ Phys.\ J.~}}

\def\cqg{{\em \/Class.\ Quant.\ Grav.~}}

\def\grg{{\em \/Gen.\ Rela.\ Gravi.~}}
\def\ijmpa{{\em \/Int.\ Journ.\ Mod.\ Phys.\ A~}}

\def\mpla{{\em \/Mod.\ Phys.\ Lett.\ A~}}

\def\jhep{{\em \/ JHEP~}}

\def\npb{{\em  \/Nucl.\ Phys.\ B~}} 
\def\phyrep{{\em  \/Phys.\ Rep.~}}

\def\plb{{\em  \/Phys.\ Lett.\ B~}}

\def\prd{{\em  \/Phys.\ Rev.\ D~}}
\def\prl{{\em  \/Phys.\ Rev.\ Lett.~}}

\begin{document}
\preprint{gr-qc/0301090}

\title[Short Title]{Temperature and entropy of Schwarzschild-de Sitter space-time}
\author{S. Shankaranarayanan\footnote{E-mail: \tt shanki@notes.uac.pt}}
\affiliation{DCTD, University of Azores, 9500 Ponta Delgada, Portugal.}

\begin{abstract}  
In the light of recent interest in {\it quantum gravity in de Sitter space},
we investigate semi-classical aspects of 4-dimensional
Schwarzschild-de Sitter space-time using the method of complex paths. 
The standard semi-classical techniques (such as Bogoliubov coefficients
and Euclidean field theory) have been useful to study quantum effects in
space-times with single horizons; however, none of these approaches
seem to work for Schwarzschild-de Sitter or, in general, for
space-times with multiple horizons. We extend the method of complex
paths to space-times with multiple horizons and obtain the spectrum of
particles produced in these space-times. We show that the temperature
of radiation in these space-times is proportional to the effective
surface gravity -- inverse harmonic sum of surface gravity of each
horizon. For the Schwarzschild-de Sitter, we apply the method of
complex paths to three different coordinate systems -- spherically
symmetric, Painleve and Lemaitre. We show that the equilibrium
temperature in Schwarzschild-de Sitter is the harmonic mean of
cosmological and event horizon temperatures. We obtain Bogoliubov
coefficients for space-times with multiple horizons by analyzing the
mode functions of the quantum fields near the horizons. We propose a
new definition of entropy for space-times with multiple horizons
analogous to the entropic definition for space-times with a single
horizon. We define entropy for these space-times to be inversely
proportional to the square of the effective surface gravity. We show that
this definition of entropy for Schwarzschild-de Sitter satisfies
the D-bound conjecture.

\end{abstract}
\maketitle 

\section{Introduction}
Over the last three decades, quantum field theory in de Sitter (dS)
space has been a subject of growing interest. In the 1970's, the
attention was due to the large symmetry group of dS space, which made
the field theory in dS space less ambiguous than, for example,
Schwarzschild space-time. In the 1980's, the focus was due to the role
it played during inflation -- accelerated expansion in the universe's
distant past. Recent attention to dS and asymptotic dS space-times is
motivated by two aspects: (i) Observations \cite{supernova} suggest
that the universe might be currently asymptotic dS and approach a pure
dS space (ii) Success of AdS/CFT correspondence \cite{maldacena} has
lead to the intense study of quantum gravity of de Sitter space
\cite{witten}. The focus has been to obtain an analogue of the AdS/CFT
correspondence in the dS space
\cite{strominger,ds-cft-earlier,witten}. For recent attempts on the
semi-classical aspects of dS and asymptotic dS space in the light
of the dS/CFT correspondence, please see
Refs. \cite{busso,parikh,medved,brickwall,odinstov-sds}.  [The authors
in Refs. \cite{parikh, medved} extended the method introduced in
Ref. \cite{parikh-old} to the \pl~coordinates of (Schwarzschild)dS.]

Even though, there has been an extensive study of the semi-classical
aspects of dS (for a recent review, see Ref. \cite{leshouches}) very
little has been understood in the case of Schwarzschild-de Sitter
(SdS) space-time. (An incomplete list of references, as regard to
semi-classical aspect, is given in
Refs. \cite{sds,paddynew,lin,odinstov}.) The fundamental difference
between SdS and dS (also Schwarzschild) space-times is the existence
of multiple horizons. SdS has two -- cosmological and event --
horizons, while dS (and Schwarzschild) space-time has {\it only} one
horizon.

Various semi-classical approaches or techniques (such as Bogoliubov
coefficient, particle detectors, effective action, Euclidean field
theory) have been used in the literature to study quantum effects in
space-times with single horizon (like dS and Schwarzschild
space-times). All the approaches conclude that the notion of
temperature (and entropy) of the space-time is associated with the
horizon. In the case of Bogoliubov coefficients, one uses the mode
functions to obtain the spectrum of particles while using Euclidean
field theory, one obtains temperature using the periodicity
arguments. Even though, these approaches work well for space-times
with single horizon, none of them work for SdS or, in general,
space-times with multiple horizons. A naive extension of these
approaches to SdS leads us to the conclusions that the SdS has two
different temperatures associated with the two horizons. Using this
extension, it has been argued \cite{medved,maeda} that the SdS will
inevitably evolve towards an empty de Sitter space indicating that SdS
may never be in thermodynamic equilibrium with a single temperature
associated with the space-time.

The above argument seems to be in contradiction with the well known
case -- Schwarzschild black-hole in thermal equilibrium with a
radiation in a bounded box. In this case, black-hole has a negative
specific heat while the radiation has a positive specific heat. The
two will be in thermal equilibrium if the box is bounded, in other
words compact. On the contrary, if the box is unbounded the black-hole
evaporates completely. The situation is identical to the case of our
interest -- black-holes in de Sitter space. The de Sitter space is
compact with no notion of spatial infinity. Besides, it has a positive
specific heat similar to the above mentioned case. The specific heat
of de Sitter space is given by \cite{paddynew,leshouches}
\beq
C_V  = \frac{\partial E}{\partial T} = 
\frac{1}{4 \pi T_{dS}^2} =  \pi l^2 = S_{dS}. 
\eeq
where $S_{dS}$ ($T_{dS}$) is the entropy (temperature) of de Sitter
space. However, the difference between de Sitter and the bounding box
is that the de Sitter has a (cosmological) horizon while the bounding
box, by construction, does not posses a horizon. The similarity of the
two systems strongly suggests that we should be able to obtain a
temperature for SdS corresponding to system in thermal equilibrium.

Given this, one would like to ask the following question: Can one
obtain a temperature for SdS which corresponds to system in thermal
equilibrium using semi-classical techniques? The purpose of this paper
is an attempt in this direction. As mentioned in earlier paragraphs,
standard quantum field theoretic techniques have not proven useful for
space-times with multiple horizons. In this paper, we extend the
method of complex paths to space-times with multiple horizons and
obtain the spectrum of particles produced in these space-times. (The
method of complex paths has proved to be useful in obtaining the
temperature associated with a quantum field propagating in a
spherically symmetric coordinate space-times with single horizon
\cite{srini,mine,mythesis,others}.) We show that the temperature of
radiation in these space-times is proportional to the effective
surface gravity -- inverse harmonic sum of surface gravity of each
horizon. In the case of Schwarzschild-de Sitter, we apply the method
of complex paths to three different coordinate systems -- spherically
symmetric, \pl~ and \lem. We show that the equilibrium temperature in
Schwarzschild-de Sitter is the harmonic mean of cosmological and event
horizon temperature. We obtain Bogoliubov coefficients for space-times
with multiple horizons by analyzing the mode functions of the quantum
fields near the horizons.

We propose a new definition of entropy for space-times with multiple
horizons analogous to the entropic definition for space-times
with single horizon. We define the entropy for these space-times to be
inversely proportional to the square of effective surface gravity. We
show that this definition of entropy for SdS satisfies the D-bound
conjecture \cite{boussobound}.

The paper is organized as follows: In section \ref{sec:gen-spher},
we discuss the general properties of spherically symmetric
space-times.  A brief description of SdS geometry is given in section
\ref{sec:class}. In section \ref{sec:multi-gen}, we apply the
method of complex paths to a general spherically symmetric space-times
and show that the equilibrium temperature in proportional to the
inverse harmonic sum of the surface gravity of each horizon. In
sections \ref{sec:spher-sds} and \ref{sec:pl-lem-sds}, we apply the
method of complex paths to three coordinate systems -- spherically
symmetric, \pl~and \lem~-- of SdS.  In section \ref{sec:entropy}, we
propose a new definition of entropy for space-times with multiple
horizons and discuss its implications for SdS. Finally, in section
\ref{sec:inter}, we discuss the results.

Throughout this paper, the metric signature we shall adopt is $(-, +,
+, +)$. We use Greek letters for $(3 + 1)$-D and lower case Latin
letters for $(1+1)$-D. Quantum field is a massless, minimally coupled
scalar field ($\Phi$).

\section{Spherically symmetric space-times}

\subsection{General analysis}
\label{sec:gen-spher}

The line-element for an interval in a spherically symmetric space-time
can be written in the following forms:
\br
ds^2&\equiv& g^{(4)}_{\mu\nu} dx^{\mu}dx^{\nu} \nonumber \\ 
\label{eq:gen-4D}
&=& g^{(2)}_{ab}dx^a dx^b + \exp[-2 \phi(x^a)] d\Omega^2 \\ &=&
- \exp[{\nu(x^0, x^1)}] (dx^0)^2\! + \exp[{\lambda(x^0, x^1)}]
(dx^1)^2\! \nonumber \\ & & + \exp[-2 \phi(x^0, x^1)] d\Omega^2\!\!
\label{eq:sphesymm}.
\er
where $d\Omega^2$ is the $2-$dimensional angular line element. As
discussed in Ref. \cite{novikov} (see also Refs. \cite{mine}), the
space-time structure of spherically symmetric space-times can be
understood {\it via} $R$ and $T$ regions. If at the given event in the
coordinate system (\ref{eq:sphesymm}), the inequality
\beq 
\exp\l[{\nu - \lambda}\r] > \l( \frac{\pa\phi}{\pa x^0}/
\frac{\pa\phi}{\pa x^1}\r)^2
\label{maincond}
\eeq
\noindent is satisfied, then the event is defined as $R$-region. [If the
above inequality is satisfied at a certain world point then by the
virtue of the continuity ($\exp(\nu - \lambda)$, $\pa \phi/\pa x^0$,
$\pa \phi/\pa x^1$ cannot be discontinuous) it is satisfied in some
neighborhood of this point.  Thus, the points in the neighborhood of
this system of coordinates satisfies the above inequality are $R$-points
and a set of them a $R$-region.] If the opposite inequality is
satisfied, the event is in a $T$-region. The definitions of $R$ and $T$
regions can be shown to be coordinate invariant.

\vspace*{0.2cm}

\noindent {\it Spherically symmetric coordinate:} Choosing the
Schwarzschild gauge, line-element (\ref{eq:sphesymm}) can be written as
\beq
ds^2 = - g(r) dt^2 + \frac{dr^2}{g(r)} + r^2 d\Omega^2 \, ,
\label{eq:gen-spher}
\eeq
where $g(r)$ is an arbitrary (continuous, differentiable) function of
$r$.  For space-times with single horizon (like Schwarzschild, dS),
$g(r)$ vanishes at one point say $r = r_0$. Near $r_0$, $g(r)$ can be
expanded as
\beq
g(r) = R(r_0) ~ (r - r_0)
\eeq
where $R(r_0)$ is twice the surface gravity ($\kappa$) of the
horizon. 

For space-times with multiple horizons (like SdS), $g(r)$ vanishes at
more than one point say $r = r_i$ where $i = 1, 2, \cdots, n$. In
general, $g(r)$ can be written in the following form
\beq
g(r) = a \frac{(r - r_1) (r - r_2) (r - r_3) \cdots (r - r_n)}{r^m} \, ,
\label{eq:def-mul-gr}
\eeq
where $a$ is a constant, $m < n$, $r_n > r_{n-1} > \cdots > r_1$ and
all $r_i$'s are assumed to be positive. Around each of these points one
can expand $g(r) = R(r_i) ~ (r - r_i)$ where $R(r_i)/2$ is the surface
gravity ($\kappa_i$) of each of these horizons.

$R$ ($T$) region in the spherically symmetric coordinate system
satisfy the inequality condition $g(r) > 0$ ($g(r) < 0$). For
space-times with multiple horizons, since $g(r)$ has multiple zeros,
there are multiple $R$ and $T$ regions.

\vspace*{0.2cm}

\noindent {\it \pl~coordinate:} In order to obtain a line element which is
regular at the horizon, we define a new time coordinate $(t_P)$ which
is related to static time coordinate $(t)$ by the relation
\beq
t = t_P \pm f(r),
\label{eq:def-tP}
\eeq
where $f$ is required to be a function of $r$ alone to ensure that the
metric remains stationary \cite{parikh,medved}. The form of $f(r)$ can
be obtained by imposing the condition that the resulting metric be
regular at the horizon. This can be realized by demanding that the
constant-time slices be flat, i. e.,
\beq
\frac{1}{g(r)} - g(r) \l[\frac{df}{dr}\r]^2 = 1 \quad
\Longrightarrow \quad \frac{df}{dr} = \frac{\sqrt{1 - g(r)}}{g(r)}.
\label{eq:def-f}
\eeq     
Substituting the expressions for $f(r)$ and $t$ in
(\ref{eq:gen-spher}), we get
\beq
ds^2 = - g(r) dt_P^2 \pm 2 \sqrt{1 - g(r)} dr dt_P + dr^2 + r^2 d\Omega^2.
\label{eq:gen-pain}
\eeq
We will refer to the above line-element as {\it \pl~coordinate}.
The above line element is a stationary -- but not a static -- system.
``$+$'' sign in the cross-term corresponds to the ingoing null 
geodesic while the ``$-$'' sign corresponds to an outgoing null 
geodesic.

For both, in-going and out-going null geodesics, the inequality
condition for the $R$ region is given by $g(r) > 0$ implying that 
the whole of space-time is doubly mapped {\it w. r. t.} spherically
symmetric coordinate system \cite{mine}. 

\vspace*{0.2cm}

\noindent {\it \lem~coordinate:} We can get rid of the cross term, in
line-element (\ref{eq:gen-pain}), by performing a transformation
of the radial coordinate ($r$). This can be achieved by demanding that
a set of curves $[t_P, r(t_P)]$ whose tangent vectors $(1, dr/d\tau)$
are orthogonal to the surface of constant \pl~ time. This gives
\beq
R \mp t_P = \int \frac{dr}{\sqrt{1 - g(r)}}.
\label{eq:rel-R-tP}
\eeq
Substituting for $r$ in (\ref{eq:gen-pain}), we get
\beq
ds^2 = - dt_P^2 + [1 - g(r)] dR^2 + r^2 d\Omega^2.
\label{eq:gen-lem}
\eeq
We will refer to the above line-element as {\it \lem~coordinate}.  The
line-element is explicitly dependent in time and test particles at
rest relative to the reference system are particles moving freely in
the given field. The coordinate system can be modeled as that natural
to a freely in-falling/out-falling observer whose velocity at radial
infinity is zero. [``$-$'' sign in equation (\ref{eq:rel-R-tP}) corresponds
to the in-falling observer while the $+$ sign in
equation (\ref{eq:rel-R-tP}) corresponds an out-falling observer.]

In terms of spherically symmetric coordinate time $(t)$, we have
\beq
R \mp t = \int \frac{dr}{g(r)} \frac{1}{\sqrt{1 - g(r)}}.
\label{eq:def-R}
\eeq                                                                           
From equation (\ref{eq:rel-R-tP}) it is can be seen that
\beq
r = {\rm function~of~}(R \mp t_P) \, . 
\eeq
For the upper sign in (\ref{eq:rel-R-tP}), defining $1 - g(r) = F(U)$
and $r^2 = G(U)$, the \lem~line element (\ref{eq:gen-lem}) in terms of
the light cone coordinates $(U, V)$ [$U \equiv R - t_P$, $V \equiv R -
t_P$] is
\br
ds^2& = &\l(\frac{F(U) - 1}{4}\r)(dV^2 + dU^2)  \nonumber \\
&+&2 \l(\frac{F(U)+ 1}{4}\r) dU dV + G(U) d\Omega^2 \, .
\label{eq:lem-light}
\er
For the line element corresponding to an in-falling observer, $R$
region is given by $F(U) > 0$. For the $T$ region, the inequality is
opposite. For an out-falling observer, the $R$ region is given by the
inequality $F(V) > 0$.

In the $T$ region there is an asymmetry in the direction of flow and
hence the $T$ regions corresponding to the in-falling and out-falling
observer will be different while the $R$ regions will be the same.
(For an elaborate discussion on this aspect, please refer to
Refs. \cite{mine}.)  Hence, the $T$ region in the \lem~coordinates is a
doubly mapped $T$ region of spherically symmetric coordinate \cite{mine}.

\begin{figure*}[!htb]
\epsfig{file=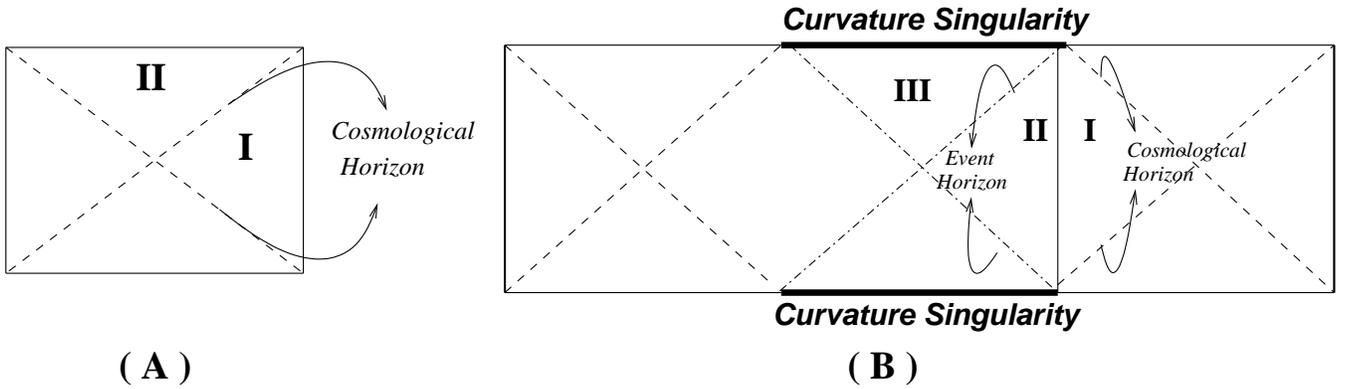,width=18cm}
\caption{Figure (A) is the Penrose diagram of the de Sitter space
where the left and right sides are identified. Spherically symmetric
coordinate system of de Sitter covers region (I) while the \pl~and
\lem~coordinate system cover regions (I) and (II). Figure (B) is the
Penrose diagram illustrating the causal structure of the Schwarzschild
de Sitter space-time. Here again, the left and right sides are
identified. Spherically symmetric coordinate system covers regions (I)
and (II). \pl~coordinate system covers regions (II) and (III).}
\end{figure*}

\subsection{Schwarzschild-de Sitter}
\label{sec:class}

The spherically symmetric coordinate of SdS space-time is given by
the line-element (\ref{eq:gen-spher}), where
\beq 
g(r) = \l(1 - \frac{2M}{r} - \frac{r^2}{l^2}\r),
\label{eq:sds-spher}
\eeq
$M$ is the mass of the black-hole and $l^2$ is related to the positive
cosmological constant. The space-time has more than one horizon if $0
< y < 1/27$ where $y = M^2/l^2$. The black hole horizon ($r_h$) and the
cosmological horizon ($r_c$) are located respectively at
\br
\label{eq:rh}
r_h &=& \frac{2 M}{\sqrt{3~y}} \cos \frac{\pi + \psi}{3} \, , \\
\label{eq:rc}
r_c &=& \frac{2 M}{\sqrt{3~y}} \cos \frac{\pi - \psi}{3} \, ,
\er
where 
\beq
\psi = \cos^{-1}\l(3 \sqrt{3 y}\r) \, .
\eeq
In the limit of $y \to 0$, we get $r_h \to 2M$ and $r_c \to l$. [Note
that $r_h < r_c$, i. e. event horizon is the smallest positive root.]
The space-time is dynamic for $r < r_h$ and $r > r_c$. In the limit $y
\to 1/27$, the two horizons -- event and cosmological -- coincide and
is the well known Nariai space-time. If $y > 1/27$, the space-time
is dynamic for all $r > 0$.

Surface gravity of these horizons are given by \cite{bousso-1996}
\br
\label{eq:surf-event}
\kappa_{h}& = & \alpha \l|\frac{M}{r_h^2} - \frac{r_h}{l^2}\r| \, ,\\
\label{eq:surf-cosmo}
\kappa_{c}& = & \alpha \l|\frac{M}{r_c^2} - \frac{r_c}{l^2}\r| \, ,
\er
where
\beq
\alpha = \frac{1}{\sqrt{1 - (27 y)^{1/3}}}\, .
\eeq
In the limit of $y \to 0$, we get $\kappa_h \to 1/(4M)$ and $\kappa_c
\to 1/l$.  The range of $\kappa_h$ and $\kappa_c$ are
\br
\label{eq:range-event}
\frac{3 \sqrt{3}}{4 l} < \kappa_h < \frac{\sqrt{3}}{l} &\Longrightarrow&
\frac{l}{\sqrt{3}} < \frac{1}{\kappa_h} < \frac{4 l}{3 \sqrt{3}} \, ,\\
\label{eq:range-cosmo}
\frac{1}{l} < \kappa_c < \frac{\sqrt{3}}{l} &\Longrightarrow&
\frac{l}{\sqrt{3}} < \frac{1}{\kappa_c} < l \, .
\er
[The above ranges are obtained by setting $y \to 0$ and $y \to
(1/27)$ in equations (\ref{eq:surf-event}, \ref{eq:surf-cosmo}).] 

Figure (IB) illustrates one version of the Penrose diagram of SdS
space-time. [For an easy comparison, we have also provided the Penrose
diagram of de Sitter space.] The static coordinate system covers
region (I) and (II) of the Penrose diagram. The boundaries of the
static region consist of (i) past and future cosmological horizons,
and (ii) past and future event horizons.

\pl~line-element of SdS is given by 
\beq 
ds^2 =  - g(r) dt_P^2 \pm 2\sqrt{\frac{2M}{r} + \frac{r^2}{l^2}} dr dt_P
+ dr^2 + r^2 d \Omega^2 \, .
\label{eq:sds-pain}
\eeq
where $t_P$ is related to $t$ by the relation 
\beq
t = t_P \pm \int dr \frac{\sqrt{2M/r + r^2/l^2}}{(1 - 2M/r - r^2/l^2} \, .
\label{eq:sds-def-tP}
\eeq

The above line element is a stationary -- but not a static -- system
and it covers regions (II) and (III) of the Penrose diagram. ``+''
sign in the above chart is suitable for studying the physical
experience of observers falling freely and radially into the
hole. [``$+$'' sign in the cross-term corresponds to the ingoing null
geodesic while the ``$-$'' sign corresponds to an outgoing null
geodesic.] For both, in-going and out-going null geodesics, the
inequality condition for the $R$ region \cite{novikov} is given by
$g(r) > 0$ implying that the whole of space-time is doubly mapped {\it
w. r. t.}  spherically symmetric coordinate system
(\ref{eq:gen-spher}).

The \lem~line-element is given by
\beq
ds^2  =  - dt_p^2 + \l(\frac{2M}{r} + \frac{r^2}{l^2}\r) dR^2 + 
r^2 d\Omega^2 \, ,
\label{eq:sds-lem} 
\eeq
where,  
{\small
\beq
2 r^{3/2} = (2M l^2) \exp\l[-\frac{3(R \pm t_p)}{2 l}\r] - 
\exp\l[\frac{3(R \pm t_p)}{2 l}\r] \, .
\eeq
}
The line-element is explicitly dependent in time and test particles at
rest relative to the reference system are particles moving freely in
the given field. $R - t_P$ in the R.H.S of the above equation
corresponds to the in-falling observer while the $R + t_P$ in the
R. H. S of the above expression corresponds to an out-falling
observer.  As like \pl, \lem~line-element also covers regions (II) and
(III) of the Penrose diagram.

For the line element corresponding to an in-falling observer, $R$
region \cite{novikov} is given by $F(U) > 0$ ($U = R -
t_P$). For the $T$ region, the inequality is opposite. For an
out-falling observer, the $R$ region is given by the inequality $F(V)
> 0$.

\section{Particle production in space-times with multiple horizons}
\label{sec:MCP}
	
In Refs. \cite{srini,mine,mythesis} the method of complex paths was
applied to space-times with single horizon. The temperature associated
with the quantum field (in the three coordinate systems -- spherically
symmetric, \pl~and \lem) in these space-times were shown to be
consistent with the temperatures obtained by using other quantum field
theoretic techniques like Bogoliubov coefficients, Euclidean field
theory, effective action. As discussed in Introduction, the standard
quantum field theoretic techniques do not work well for space-times
with multiple horizons. In this section, we extend the method of
complex paths to space-times with multiple horizons and in particular
to Schwarzschild-de Sitter space-time. We show that the effective
temperature for the quantum fields in these space-times is
proportional to the inverse harmonic sum of surface gravity at
each horizon.

\subsection{General spherically symmetric space-time}
\label{sec:multi-gen}

In this subsection, we obtain the effective temperature in a general
spherically symmetric space-time with multiple horizons. [The result,
in this subsection, is due to Ref. \cite{pad-me} and is summarized here
for completeness.]

Consider a quantum field propagating in a general spherically
symmetric space-time (\ref{eq:gen-spher}) with multiple horizons. The
wave equation is given by
\br
\pa_{\mu}\l(\sqrt{-g}g^{\mu\nu}\pa_{\nu}\r)\Phi =  0 \, ,& & \nonumber \\
\frac{r^2}{g(r)}\frac{\pa^2 \Psi}{\pa t^2} - \frac{\pa}{\pa
r}\l(r^2 g(r) \frac{\pa \Psi}{\pa r}\r) + L^2 \Psi = 0\, . & &
\label{eq:wav-gen-spher}
\er
where $\Phi(x^{\mu}) = \Psi(t,,r) Y_{lm}(\theta,\phi)$. The
semi-classical wave-functions satisfying the above are obtained by
making the ansatz
\beq
\Psi(r,t) = \exp\l[\frac{i}{\hbar} S(r,t)\r]\, ,
\label{eq:semi-class}
\eeq
where $S$ is a functional which will be expanded in powers of $\hbar$.
Expanding $S$ in a power series of $\hbar$:
\beq
S(r,t) = S_0(r,t) + \hbar S_1(r,t) + \hbar^2 S_2(r,t)+\dots
\,,
\eeq
and substituting the resulting expression in
equation (\ref{eq:wav-gen-spher}) (neglecting terms of order $\hbar$ or
higher), we have
\beq
\label{eq:gen-HJ}
-\frac{1}{g(r)}\left(\frac{\d S_0}{\d t}\right)^2 + g(r)
\left(\frac{\d S_0}{\d r}\right)^2 + \frac{L^2}{r^2} =0 \,.  
\eeq
This is simply the Hamilton-Jacobi equation satisfied by a massless
particle moving in a general spherically symmetric coordinate. Making
the ansatz $S_0 = - E t + B(r)$, we get
\br 
B(r)&= &\pm \int^r \frac{dr}{g(r)} \sqrt{E^2 - (L^2/r^2) g(r)}
\nonumber \\ &=& \pm \frac{E}{a} \int^r dr~\frac{r^m}{(r - r_1) (r -
r_2) \cdots (r - r_n)}
\label{eq:S0-gen-sp}
\er
[The last expression is obtained by setting $L = 0$ by noticing that
near the horizon, the presence of the $L^2$ term can be neglected. It
can also be noted that the above expression is valid for an arbitrary
dimension and hence the following result is true for a spherically
symmetric space-time in an arbitrary dimension.]  Rewriting R. H. S in
the above expression, by partial fractions, {\it i. e.}
\beq
\frac{1}{a}\frac{r^m}{(r - r_1) (r - r_2) \cdots (r - r_n)} = \sum_{i = 1}^{n} 
\frac{A_i}{(r - r_i)}
\label{eq:partial-gen}
\eeq
where 
\br
A_i& = & \frac{{r_i}^m}{a}\prod_{j = 1}^{n}\frac{1}{(r_i - r_j)} \quad  j \neq i 
\nonumber \\
&=& \frac{1}{(m - 1) \kappa_i}
\label{eq:Ai-gen}
\er
and $\kappa_i$ is the surface gravity of each of the
horizons. Substituting the above expressions in equation
(\ref{eq:S0-gen-sp}), we get,
\beq
B(r_s, r_f)= \pm \frac{E}{m - 1} \sum_{i = 1}^{n} \frac{1}{\kappa_i} 
\int_{r_s}^{r_f}
\frac{dr}{r - r_i} 
\eeq
The sign ambiguity is related to the {\it outgoing} ($\pa S_0/\pa r >
0$) or {\it ingoing} ($\pa S_0/\pa r < 0$) nature of the
particle. Unlike space-times with single horizon
\cite{srini,mine}, space-times with multiple horizons have many
possible ways of choosing initial ($r_1$) and final ($r_2$)
points. However, we will consider a particular situation of the
particle crossing all the horizons, {\it i. e.} $r_s < r_1$ and $r_f >
r_n$. [This result can be used to obtain other possible cases.]

For an outgoing particle ($\pa S_0/\pa r > 0$) w.r.t the horizon
($r_1$), the contribution to $S_0$ is
%
\br
S_0({\rm emission}) &=& - \frac{E }{(m - 1)\kappa_1}
\int_{r_1 - \e}^{r_1 + \e } \frac{dr}{r - r_1} \nonumber \\
&-& \frac{E}{(m - 1) \kappa_2}\int_{r_2 + \e}^{r_2 - \e} 
\frac{dr}{r - r_2} - \cdots \nonumber \\
&-& \frac{E}{(m - 1) \kappa_n}\int_{r_n + \e}^
{r_n - \e}\frac{dr}{r - r_n} + {\rm real~part}\nonumber \\
&=& i \frac{\pi E }{(m - 1) \kappa_{\rm eff}} + {\rm real~part}
\label{eq:s0emi-gen}
\er
%
where 
\beq
\frac{1}{\kappa_{\rm eff}} = \sum_{i = 1}^{n}\frac{1}{\kappa_i} \, ,
\label{eq:keff-gen}
\eeq
and the ``$-$'' sign in-front of the integrals correspond to the
condition that $\pa S_0/\pa r > 0$ at $r = r_s < r_1$ and $\pa S_0/\pa
r < 0$ at $r = r_f > r_n$ and the integrals are evaluated by taking
the contours to lie in the upper complex plane. It should be noted
that the above expression can also be considered as the action for the
ingoing particles w.r.t the horizon $r_n$. [Note that we have assumed
all $r_i$'s to be positive, and hence $S_0$ will have {\it imaginary}
contribution from all of them. For SdS, which we consider in detail,
one of the roots of $g(r)$ is negative and will not contribute to
$S_0$.]

Using the above procedure, the action for an ingoing particle ($\pa
S_0/\pa r < 0$) w.r.t the horizon $r_1$ is given by
%
\br
S_0({\rm absorption})&=& - \frac{E }{(m - 1)\kappa_1}
\int_{r_1 + \e}^{r_1 - \e } \frac{dr}{r - r_1} \nonumber \\
&-& \frac{E}{(m - 1) \kappa_2}\int_{r_2 - \e}^{r_2 + \e} \frac{dr}{r -
r_2}- \cdots   \nonumber \\
&-& \frac{E}{(m - 1) \kappa_n}\int_{r_n - \e}^
{r_n + \e} \frac{dr}{r - r_n} + {\rm real~part}\nonumber \\
&=& - i \frac{\pi E}{(m - 1) \kappa_{\rm eff}} + {\rm real~part}
\label{eq:s0abs-gen}
\er
%
the last expression is obtained by evaluating the integrals with the
contours taken in the lower half-plane. Taking the modulus square of
equations (\ref{eq:s0emi-gen}, \ref{eq:s0abs-gen}), we get
\beq
P[{\rm emission}] = \exp\l[-\frac{4 \pi E }{(m - 1)\kappa_{\rm eff}}\r]
P[{\rm absorption}] \, .
\eeq
The exponential dependence on the energy allows one to give a {\it
thermal} interpretation for the above result. The temperature of the
emission spectrum is given by
\beq
\beta^{-1} = \frac{(m - 1)}{4 \pi} \kappa_{\rm eff}
\, .
\label{eq:temp-gen}
\eeq
The above result shows that the spectrum of particles created in a
spherically symmetric space-time with multiple horizons is thermal and
the temperature is proportional to the inverse harmonic sum of surface
gravity of each horizon. The above result can be interpreted as a
particle propagating from inside the horizon ($r_1$) to outside the
horizon ($r_n$) picks up $\exp(-\beta_i E)$ at each horizon, resulting
in $\exp[- \beta E)$. [Note that, we have neglected the
back-scattering of the particles from the horizon.] The above analysis
can be performed to \pl~ and \lem~coordinates for these
multiple-horizons space-times as discussed in Appendix
(\ref{app:pp-single}). In the next section, we explicitly demonstrate
the above result for three different coordinate systems of SdS
space-time.

\subsection{Spherically Symmetric SdS coordinate}
\label{sec:spher-sds}

For a quantum field propagating in spherically symmetric coordinate
system of SdS space-time, the field equation is given by
(\ref{eq:wav-gen-spher}) where $g(r)$ is given by
(\ref{eq:sds-spher}). In the previous section we have assumed that all
the roots of $g(r)$ are positive and hence $S_0$ has contributions
from all $r_i$'s. However, for SdS, one of the roots of $g(r)$ is
negative and the results in the previous section does not follow
automatically.

Making the usual semi-classical ansatz for
$\Psi$, and expanding $S$ in the powers of $\hbar$, we get
\br
B(r)&= &\pm  \int^r \frac{dr}{g(r)} \sqrt{E^2 - L^2/r^2 g(r)} \nonumber \\
& = & \pm E \int^r \frac{dr}{1 - 2M/r - r^2/l^2} \, ,
\label{eq:S0-sds-sp}
\er
where the last expression is obtained by setting $L = 0$. After a bit
of lengthy algebra, it can be shown that
\br
\l(1 - \frac{2M}{r} - \frac{r^2}{l^2}\r)^{-1}\!\!\!&=&\frac{\alpha}{2 \kappa_h (r - r_h)}
- \frac{\alpha}{2 \kappa_c (r - r_c)} \nonumber \\
\!\!\!&-& \frac{\alpha}{2} \l(\frac{1}{\kappa_h} - \frac{1}{\kappa_c}\r) 
\frac{1}{r + r_c + r_h} \, ,
\er
where $\kappa_h, \kappa_c$ are given by expressions
(\ref{eq:surf-event}, \ref{eq:surf-cosmo}) and $r_h, r_c$ are given by
(\ref{eq:rh}, \ref{eq:rc}). Substituting the above expression in
equation (\ref{eq:S0-sds-sp}), we get,
\br
B(r)&=& \pm \frac{E \alpha}{2 \kappa_h} \int^r \frac{dr}{r - r_h} \mp 
\frac{E\alpha}{2 \kappa_c} \int^r \frac{dr}{r - r_c} \nonumber \\
&\mp& \frac{E\alpha }{2} \l(\frac{1}{\kappa_h} - \frac{1}{\kappa_c}\r) 
\int^r \frac{dr}{r + r_c + r_h} 
\label{eq:sds-B}
\er
The sign ambiguity is related to the {\it outgoing} ($\pa S_0/\pa r
> 0$) or {\it ingoing} ($\pa S_0/\pa r < 0$) nature of the
particle. Unlike space-times with single horizon
\cite{srini,mine}, SdS has many possible ways of choosing the
initial ($r_1$) and the final ($r_2$) points. Following scenarios will
be of our interest:
\begin{enumerate}
\item $r_1 < r_h$ and $r_2 > r_c$: The first two integrals in the
R. H. S of equation (\ref{eq:sds-B}) does not exist.
\item $r_1 < r_h$ and $r_2 < r_c$: The first integral in the R. H. S
of above expression does not exist since the denominator vanishes at
$r = r_h$.
\item $r_1 > r_h$ and $r_2 > r_c$: The second integral in the R. H. S
of above expression does not exist since the denominator vanishes at
$r = r_h$.  
\end{enumerate}

In what follows, we shall obtain the spectrum of particles produced in 
the first scenario. [The other two cases follows automatically from this
discussion.]

In the first case, both the points are outside the region $(r_h,
r_c)$. Hence, the contribution to the probability amplitude for
emission and absorption will be due to both the horizons -- event and
cosmological. For an outgoing particle ($\pa S_0/\pa r > 0$) w.r.t the
event horizon ($r_h$), the contribution to $S_0$ is
\br
S_0({\rm emission})&=& - \frac{E\alpha }{2 \kappa_h}
\int_{r_h - \e}^{r_h + \e } \frac{dr}{r - r_h} \nonumber \\
&-& \frac{E\alpha }{2 \kappa_c}\int_{r_c + \e}^{r_c - \e} \frac{dr}{r - r_c} 
+ {\rm real~part} \\
&=& i \frac{\pi E\alpha }{2 \kappa_{\rm eff}} + {\rm real~part} \, ,
\er
where 
\beq
\kappa_{\rm eff} = \l[\frac{1}{\kappa_h} + \frac{1}{\kappa_c}\r]^{-1}\, ,
\label{eq:kappa-def}
\eeq
and the ``$-$'' sign in-front of the integrals correspond to the
condition that $\pa S_0/\pa r > 0$ at $r = r_1 < r_h$ and $\pa S_0/\pa
r < 0$ at $r = r_2 > r_c$ and the integrals are evaluated by taking
the contours to lie in the upper complex plane. It should be noted
that the above expression can also be considered as the action for
the ingoing particles w.r.t the cosmological horizon ($r_h$).

Using the above procedure, the action for an ingoing particle ($\pa
S_0/\pa r < 0$) w.r.t the event horizon ($r_c$) is
\br
S_0({\rm absorption})&=& - \frac{E\alpha}{2 \kappa_h}
\int_{r_h + \e}^{r_h - \e } \frac{dr}{r - r_h} \nonumber \\
&-& \frac{E\alpha}{2 \kappa_c}\int_{r_c - \e}^{r_c + \e} \frac{dr}{r - r_c} 
+ {\rm real~part} \\
&=& - i \frac{\pi E\alpha}{2 \kappa_{\rm eff}} + {\rm real~part} \, ,
\er
where, the integrals are evaluated by taking the contour in the lower
half-plane. Taking the modulus square, we get
\beq
P[{\rm emission}] = \exp\l[-\frac{2 \pi E\alpha}{\kappa_{\rm eff}}\r]
P[{\rm absorption}] \, . 
\eeq
The exponential dependence on the energy allows one to give a {\it
thermal} interpretation for the above result. The temperature of the
emission spectrum is given by
\beq 
\beta^{-1} = \frac{1}{2\pi\alpha} \frac{\kappa_h
\kappa_c}{\kappa_h + \kappa_c} \, .
\label{eq:temp1-sds}
\eeq

Following features are note-worthy regarding this result:

(i) The equilibrium temperature in SdS is inversely proportional to
the harmonic mean of the two horizons. [The range of $\kappa_{\rm eff}$ 
is $(1/l) < \kappa_{\rm eff} < (\sqrt{3}{l})$.] 

(ii) In this case, radiation is propagating inwards from the
cosmological horizon and outwards from the black-hole horizon. [We
have neglected the scattering of the radiation from these horizons.]
Thus, there is a constant flux of radiation flowing between these
horizons and a static observer is in a thermal bath of radiation with
the above temperature. [This result has been obtained in Ref. \cite{lin} 
from Euclidean quantum gravity point-of-view.]

(ii) In the limit of $1/l \to 0$, it reproduces Hawking temperature for
a Schwarzschild black-hole in asymptotically flat space-time.

(iii) In the limit of $M \to 0$, this reproduces Gibbons-Hawking 
temperature for a vacuum de Sitter space-time. 

(iv) In the Nariai limit ($y \to 1/27$), the event and cosmological
horizons coincide. Thus, the results obtained for space-times with
multiple-horizons can not be applied. In the Nariai limit, using the
analysis performed in Ref. \cite{srini} and Appendix
\ref{app:pp-single}, we get \cite{bousso-1996}
\beq
\beta_{\rm Nariai}^{\rm -1} = \frac{\sqrt{3}}{2\pi l}.
\eeq

\subsection{\pl~ and \lem~ SdS coordinate}
\label{sec:pl-lem-sds}

Let us now consider scalar field propagating in \pl~coordinate system
of SdS space-time. Retracing the steps as discussed in appendix
(\ref{app:pp-single}), we get (for $l = 0$),
\beq
B(r) = E \int^r dr \frac{\sqrt{2M/r + r^2/l^2} \pm 1}{1 - 2M/r - r^2/l^2}
\label{eq:S0-sds-pain}
\eeq
As discussed in section (\ref{sec:class}), \pl~coordinate covers
regions II and III. In the \pl~coordinate system, as opposed to
spherically symmetric coordinate, there are only two possible ways of
choosing the initial ($r_1$) and final ($r_2$) points: (i) both the
points are on one side of $r_h$ and (ii) the two points lie on the
opposite sides of the event horizon. 

For an outgoing particle ($\pa S_0/\pa r > 0$) w.r.t the event horizon
($r_c$), the contribution to $S_0$ is
\br 
S_0[{\rm emission}] & =& - \frac{E\alpha}{\kappa_h} \int_{r_h - \e}^{r_h
+ \e } \frac{dr}{r - r_h} + {\rm real~terms} \nonumber \\ &=& i
\frac{\pi E\alpha}{\kappa_h}  + {\rm real~terms}
\er
In order to obtain the action for absorption of particles
corresponding to the ingoing particles [$(\pa S_0/\pa x) < 0$], we
have to repeat the above calculation for the lower sign of the metric
given in equation~(\ref{eq:sds-pain}). In this case, it is easy to see
that the only singular solution corresponds to in-going particles and
we get
\beq
S_0[{\rm absorption}] = - \frac{\pi E\alpha}{\kappa_h}  + {\rm real~terms}
\eeq
Constructing the semi-classical propagator in the usual manner and
taking the modulus square we obtain the probability. From the earlier
discussion, we know that in calculating the probability of
absorption/emission there is an extra contribution to the probability is 
from four sets of complex paths satisfying the semi-classical ansatz. 
Taking this into account, we get 
\beq
\beta^{-1} = \frac{\kappa_h}{2\pi\alpha} \, .
\eeq
In the case of \lem~coordinates, repeating the
above procedure and it is easy to show that the temperature of the
radiation is same as that obtained in the case of \pl~coordinates and
case II of Spherically symmetric coordinates. The \pl~and \lem~observers
will be in thermal equilibrium with the above temperature.

\section{Bogoliubov coefficients for space-times with multiple horizons}
\label{sec:bogo}

In this section, we extend the analysis of appendix (\ref{app:bogo})
and obtain Bogoliubov coefficients for space-times with multiple
horizons. The equation of the motion for the scalar field in the
spherically symmetric coordinate is given by equation
(\ref{eq:wav-gen-spher}).  Near each of the horizons, the solution to
the wave equation is given by
\beq 
\Psi = C^i_1 \exp(-iE t)~ (x_i)^{i E/R(r_i)},
\label{eq:sol-many}
\eeq
where $C^i_1$ is an arbitrary constant. $\Psi^*$ is also a solution to
the scalar field equation (\ref{eq:wav-gen-spher}). For each horizon,
we can write the two sets of modes as
\br
\Psi(x_i < 0)& = & C^i_1 \psi(-x_i) + C^i_2 \psi^*(-x_i)\, , \nonumber \\
\Psi(x_i > 0)& = & C^i_{\alpha} \psi (x_i) + C^i_{\beta} \psi^*(x_i) \, , 
\label{eq:bogol-main-many}
\er
where $C^i_1$, $C^i_2$, $C^i_{\alpha}$ and $C^i_{\beta}$ are constants
to be determined. We obtain the relation between the different
constants similar to the single horizon case. They are given by
\br
\label{eq:bogol1-many}
C^i_{\alpha}& = & C^i_1 \exp[\pi E/R(r_i)] \\ 
C^i_{\beta}& = & C^i_2 \exp[-\pi E/R(r_i)]\, .
\label{eq:bogol2-many}
\er
This leaves us with identifying relation between the two constants
($C^i_1$, $C^i_2$) to the left of the horizon $r_i$. For space-times
with single horizon , we assumed that the net current to the left of
the horizon is zero. However, for space-times with multiple horizon,
the above assumption is not valid.

In order to see this explicitly, let us consider quantum field
propagating in spherically symmetric coordinate of SdS which has two
horizons -- event and cosmological. Let $r_i$ be the cosmological
horizon. To the left of the cosmological horizon, the net-current is
non-zero due to the presence of event horizon which produces thermal
radiation. [Here we are assuming that the back-scattering of the
particles from the event-horizons can be neglected.]  In this case,
$C^i_1$ and $C^i_2$ are related by
\beq
|C^i_1|^2 =  \exp(-2 \pi \alpha E/\kappa_h) |C^i_2|^2 \,.
\label{eq:bogo-cond}
\eeq
This gives,
\beq
\l|{C^i_{\beta} \over C^i_{\alpha}}\r|^2 = \exp[-2 \pi \alpha E/\kappa_{\rm eff}].
\label{eq:bogol3-many}
\eeq
Interpreting $|C^i_{\alpha}|^2 \equiv |\alpha_E^i|^2$ and 
$|C^i_{\beta}|^2 \equiv |\beta_E^i|^2$ as Bogoliubov coefficients and 
using the unitary condition, we have 
\br
\l|\alpha^i_E\r|^2 & = & {1 \over 1 -\exp[-2\pi \alpha E/\kappa_{\rm eff}] } \nonumber \\
\l|\beta^i_E\r|^2  & = & {1 \over \exp[2\pi \alpha E/\kappa_{\rm eff}] - 1}.
\label{eq:bogol4-many}
\er
Following features are noteworthy regarding this result:
\begin{enumerate}
\item $|\beta^i_E|^2$ is a Planck spectrum with a temperature given by
equation (\ref{eq:temp1-sds}). 
\item The Bogoliubov coefficients at each of the horizon are same
implying that the space-time is in thermal equilibrium with
temperature proportional to $\kappa_{\rm eff}$.
\item The condition in equation (\ref{eq:bogo-cond}) has been crucial in
obtaining the result. The condition implies that a particle
propagating from inside the horizon ($r_h$) to outside the horizon
($r_c$) picks up $\exp[-\beta_h E]$ and $\exp[-\beta_c E]$ resulting
in $\exp[- \beta E)$.
\item In the Nariai limit, the condition (\ref{eq:bogo-cond}) is not
valid and we need to impose the condition (\ref{eq:bogolcond}) (which
implies that the net current to the left of the horizon is zero). In
this case, the Bogoliubov coefficients are given by equations
(\ref{eq:bogol4}) where $\kappa = l/\sqrt{3}$.
\end{enumerate}

\section{Entropy of SdS}
\label{sec:entropy}

In this section, we propose a new definition of entropy for space-times
with multiple horizons analogous to the entropic definition for
space-times with single horizon and discuss the implications in the
context of SdS.

For space-times with (compact) single horizon, the entropy is 
given by
\beq
S({\rm single~ horizon}) =  \frac{A}{4} = c \frac{\pi}{\kappa^{2}} \, .
\eeq
where $c = 1, 1/4$ for dS and Schwarzschild respectively. Assuming
that the relation between $S$ and $\kappa$ holds true for space-times
with multiple horizons, the entropy for space-times with multiple
horizons can be written as 
\beq
S({\rm multiple~ horizons}) = \frac{\pi}{\kappa_{\rm eff}^2} \, ,
\eeq
where $\kappa_{\rm eff}$ is given by equation
(\ref{eq:keff-gen}). Following features are noteworthy regarding this
expression:

(i) The above expression can be treated as an entropy associated with
a single (effective) horizon for space-times with multiple
horizon. For SdS, this corresponds to the entropy of the system in 
thermal equilibrium with the temperature (\ref{eq:temp1-sds}).

(ii) In the literature, entropy of space-times with multiple horizons
is defined as the sum of entropies of individual horizons. Using our
definition, the entropy will have extra non-zero contributions. In the 
case of SdS, we have
\beq
S_{SdS} = \frac{\pi}{\kappa_h^2} + \frac{\pi}{\kappa_c^2} + \frac{2 \pi}{
\kappa_c \kappa_h}. 
\label{eq:entr-sds}
\eeq
If we set
\beq
\frac{\pi}{\kappa_h^2} = S_h; \frac{\pi}{\kappa_c^2} = S_c
\eeq
Then, the expression for entropy of SdS can be written as 
\beq
S_{SdS} = S_h + S_c + 2 \sqrt{S_c S_h} \, ,
\label{eq:sds-ent}
\eeq
where $S_h, S_c$ are the entropies of event and cosmological horizons,
respectively. 

(iii) The entropy of SdS obtained in equation (\ref{eq:entr-sds})
satisfies the D-bound conjecture \cite{boussobound}.

Following the discussion in section (\ref{sec:class}), we know that
the range of $\kappa_{\rm eff}$ is $(1/l, \sqrt{3}/l)$.  [The range of
$1/\kappa_{\rm eff}$ is $(l/\sqrt{3}, l)$.] Substituting the range of 
$\kappa_{\rm eff}$ in the expression for entropy, we get
\beq
S_{SdS} < \pi l^2 \equiv S_{dS} \, .
\label{eq:ineq}
\eeq

(iv) The form of the entropy in equation (\ref{eq:sds-ent}) is
similar to equation (5.1) of Ref. \cite{maeda}.  In Ref. \cite{maeda},
the authors have shown that the final values of entropies of the event
($S_B$) and cosmological ($S_C$) horizon satisfy the inequality
\beq
S_B + S_C + \sqrt{S_B S_C} \leq \pi l^2.
\label{eq:maeda-ine}
\eeq 
In our case, the inequality (\ref{eq:ineq}) is satisfied for all
values of entropies of the event and cosmological horizons.  The
D-bound we have obtained in equation (\ref{eq:ineq}) is a stronger
inequality than that of (\ref{eq:maeda-ine}).

\section{Conclusions}
\label{sec:inter}

In this paper, we have studied the spectrum of created particles in
SdS space-time for a linear, massless scalar field using the method of
complex paths. We have shown that it is possible to obtain a
temperature for SdS which corresponds to system in thermal
equilibrium. The equilibrium temperature is the harmonic mean of the
event and cosmological horizon temperature. We have also obtained the
same result by calculating the Bogoliubov coefficients using the
wave-modes near the horizons.

It has been assumed in the literature that the entropy of space-times
with multiple horizons is the sum of entropies of individual horizons.
In this paper, we have proposed a new definition of entropy for
space-times with multiple horizons analogous to the entropic
definition of entropy for space-times with single horizon. We have
defined the entropy for these space-times to be inversely proportional
to the square of the effective surface gravity. Using this definition,
we have shown that the entropy of SdS satisfies D-bound conjecture
\cite{boussobound}. In Ref. \cite{mann}, the authors have proposed a
definition for entropy outside the cosmological horizon in SdS which
satisfies D-bound. It is interesting to look for the connection
between the two proposed entropies.

The above result brings attention to the following interesting questions:
\begin{enumerate}
\item For space-times with single horizon, there is a unique way of
obtaining a global coordinate system. For the space-times with
multiple horizons, it has not been possible to obtain a global
coordinate system. Is it possible to obtain a global coordinate system
whose Euclidean metric has periodicity proportional to $\kappa_{\rm eff}$?
\item Can we obtain stress tensors for SdS which corresponds to the
thermal state with the temperature proportional to $\kappa_{\rm eff}$?
\item What kind of quantum vacuum state will this correspond to?
\end{enumerate}
We hope to return to study some of these problems in the near future.
\begin{acknowledgments}
\noindent     
The author would like to thank T. Padmanabhan and A. J. M. Medved for
useful discussions.  The author gratefully acknowledges support from
Funda\c c\~ao para a Ci\^encia e a Tecnologia (Portugal) under the
grant SFRH/BI/9622/2002.
\end{acknowledgments}

\appendix

\section{}
\label{app:pp-single}

In this appendix, we generalize the results of Ref. \cite{mine} to two
non-singular coordinate systems -- \pl~ and \lem~ -- of a a general
spherically symmetric space-time (with single horizon). We show that
the temperature associated with the radiation in the these two
coordinate systems, {\it for space-times with single horizon}, are
same as the spherically symmetric coordinate system. These results can
be extended to space-times with multiple horizons where $g(r)$ is
given by equation (\ref{eq:def-mul-gr}).

In the context of method of complex paths, these two coordinate
systems possess interesting feature of double mapping of the paths.
In the following, we briefly describe the multiple of mapping of
space-time and measure of these paths. In Refs. \cite{mine}, it was
argued that, the family of complex paths, used to calculate the
emission/absorption probability, takes into account of all paths
irrespective of the multiple mapping of that part/whole of the
space-time.

Let us consider coordinate system like \pl~where whole of the
space-time is doubly mapped {\it w.r.t} the spherically symmetric
coordinate. The space-time has two distinct $R$ and $T$
regions. Hence, the complex paths will have {\it equal} contributions
from the both of these, which do not any point in common. Hence, the
contribution to the amplitude of emission/absorption by these two
paths will be mutually exclusive.

In the case of coordinate systems where the part of the space-time is
doubly mapped {\it w. r. t.} the spherically symmetric coordinate (as
in the case of \lem~coordinate), it is always possible to find one
point that is common to the paths contributing to
absorption/emission. Hence, these paths are not mutually
exclusive. These paths, on the other hand, will be mutually exclusive
when one considers the probability amplitude -- which is the important
quantity in our approach. Hence, the action we obtain by regularizing
the singularity and the resulting probability amplitude, for
absorption/emission, will have equal contributions from both these
paths.

\subsection{\pl~ coordinate system}
\label{sec:gen-pain}

Let us consider a quantum field propagating in the \pl~line-element
(\ref{eq:gen-pain}) with the ``$+$'' signature in the cross term. The
equation for a scalar field propagating in general \pl~coordinate is
given by
\br
\!\!& & r^2 \frac{\pa^2 \Psi}{\pa t_P^2} + 2 r^2 \sqrt{1 - g(r)} \frac{\pa^2 \Psi}
{\pa r \pa t_P} +\frac{d}{dr}[r^2 g(r)] \frac{\pa \Psi}{\pa r} \\
\label{eq:eom-pain}
&&- r^2 g(r) \frac{\pa^2 \Psi}{\pa r^2} 
+\frac{d}{dr}\l[r^2 \sqrt{1 - g(r)}\r] \frac{\pa \Psi}{\pa t_P}
= - L^2\Psi \, ,\nonumber
\er
where $\Phi(x^{\mu}) = \Psi(t_P,r) Y_{lm}(\theta,\phi)$.  Making the
standard semi-classical ansatz for $\Psi$ and expanding $S$ in powers
of $\hbar$ as discussed in section (\ref{sec:multi-gen}), we get, to
the lowest order,
\br
- \l(\frac{\pa S_0}{\pa t_P}\r)^2 + 2 \sqrt{1 - g(r)} \l(\frac{\pa
S_0}{\pa r}\r) \l(\frac{\pa S_0}{\pa t_P} \r) & &\nonumber \\ 
+ g(r)\l(\frac{\pa S_0}{\pa r}\r)^2 + \frac{L^2}{r^2}& = & 0 \, .
\label{eq:HJ-pain}
\er
The above equation is the Hamilton-Jacobi equation of massless particle
propagating in general \pl~line-element (\ref{eq:gen-pain}).  Substituting
the ansatz, $S_0 = -E t_P + B(r)$ in the above expression, we get 
\br
\frac{dB}{dr}&=& - \frac{E \sqrt{1 - g(r)}}{g(r)}  \\
&\pm& \frac{\sqrt{4 E^2 [1 - g(r)] - 4 g(r)[E^2 - L^2/r^2]}}{g(r)} \nonumber
\er
It is easy to see that near the horizon, the presence of the $L^2$ term
can be neglected since it is multiplied by $g(r)$. Thus for $L = 0$, we 
have,
\br
B(r) & = & E \int^r dr \frac{\sqrt{1 - g(r)} \pm 1}{g(r)} \nonumber \\
& = & E \int^r dr \frac{\sqrt{1 - g(r)}}{g(r)} \pm E \int^r \frac{dr}{g(r)} 
\, .
\er
For space-times with single horizon, one can expand $g(r)$ around the
horizon as $g(r) = R(r_0) (r - r_0)$. The numerator in the first
integral (in the R. H. S) of the above expression can be approximated
to unity near the horizon. Noticing that the denominator is singular
at $r = r_0$ only for the positive sign, we get the action for the
outgoing particle as
\br
S_0[{\rm emission}] & = & - 2 E \int_{r_0
- \e}^{r_0 + \e} \frac{dr}{g(r)} + {\rm real~part}\nonumber \\
&=& + \frac{2 i \pi E}{R(r_0)} + {\rm real~part} \, .
\label{eq:emi-pain}
\er

In order to obtain the action for absorption of particles
corresponding to the ingoing particles [$(\pa S_0/\pa r) < 0$], we
have to repeat the above calculation for the metric
(\ref{eq:gen-pain}) with ``$-$'' sign in the cross term. In this
case, it is easy to see that the only singular solution corresponds to
in-going particles and so
\beq
S_0[{\rm absorption}] = - \frac{2 i \pi E}{R(r_0)} + {\rm real~part}
\label{eq:abs-pain}
\eeq
Constructing the semi-classical propagator in the usual manner and
taking the modulus square we obtain the probability. Extending the
double mapping of the paths to the generalized \pl~coordinates and
squaring the modulus to get the probability, we get the temperature
associated with the quantum fields propagating in generalized
\pl~coordinates to be
\beq
\beta^{-1} = \frac{R(r_0)}{4\pi} =  \frac{\kappa(r_0)}{2 \pi}\, .  
\eeq 
For dS, $g(r) = (1 - r^2/l^2)$, $\kappa = (1/l)$ and hence, the
temperature associated with the radiation is $1/(2 \pi l)$.

\subsection{\lem~ coordinate system}
\label{sec:gen-lem}

Consider a quantum field propagating in the general spherically
symmetric space-times described by the \lem~line-element
(\ref{eq:gen-lem}).  The scalar field equation [for the ``$-$'' sign
in the expression (\ref{eq:rel-R-tP})] is given by
\br
- \frac{\pa}{\pa t_P}\l[r^2 \sqrt{1 - g(r)}\frac{\pa \Psi}{\pa t_P}\r] 
+ \frac{\pa}{\pa R}\l[\frac{r^2}{\sqrt{1 - g(r)}} \frac{\pa \Psi}{\pa R}\r] 
& &  \\
- \sqrt{1 - g(r)} L^2 \Psi &=& 0, \nonumber
\er
where $\Phi(x^{\mu}) = \Psi(t_P,R) Y_{lm}(\theta,\phi)$.  Noting from
equation (\ref{eq:rel-R-tP}) that $r$ is related to $R - t_P \equiv
U$, the above equation rewritten in terms of the light-cone
coordinates $(U, V)$ translates into
\br
& &\frac{\pa}{\pa V}\l[(F(U) - 1) G(U) \frac{\pa \Psi}{\pa V}\r] 
+ \frac{\pa}{\pa U}\l[(F(U) - 1) G(U) \frac{\pa \Psi}{\pa U}\r]   
\nonumber \\
& + & \frac{\pa}{\pa V}\l[(F(U) + 1) G(U) \frac{\pa \Psi}{\pa V}\r]
+ \frac{\pa}{\pa U}\l[(F(U) + 1) G(U) \frac{\pa \Psi}{\pa V}\r]
\nonumber \\
&- & 2 \frac{\sqrt{F(U)}}{G(U)} L^2 \Psi = 0 \, .
\er
Making the standard semi-classical ansatz for $\Psi$ and expanding $S$
in powers of $\hbar$, we get, to the lowest order of $S_0$,
\br 
\l(\frac{\pa S_0}{\pa U}\r)^2 + \l(\frac{\pa S_0}{\pa V}\r)^2 + 2
\l(\frac{1 + F(U)}{1 - F(U)}\r) \frac{\pa S_0}{\pa U}\frac{\pa
S_0}{\pa V} & & \\ 
- \l(\frac{L^2}{1 - F(U)}\r)& =& 0 \, .
\nonumber 
\er
The above equation is the Hamilton-Jacobi equation of massive particle
propagating in the \lem~line-element (\ref{eq:gen-lem}). Substituting
the ansatz, $S_0 = -E V + B(U)$ in the above expression, we get (for
$L = 0$)
\beq 
B(U) =  E \int^U dU \, \frac{1 + F(U) \pm 2 \sqrt{F(U)}}{1 -
F(U)} \, .
\label{eq:act-lem}
\eeq
[Near the horizon the contribution due to $L$ can be neglected since
it is multiplied by $g(r)$.]  Notice that the denominator is singular
at $U=1$ only for the positive sign. Our interest in this exercise is
to obtain the principal part of the action near the horizon $r =
r_0$. The transformation relation (\ref{eq:rel-R-tP}) near the horizon
translates into
\beq
U = - \frac{\sqrt{1 - R(r_0) (r - r_0)}}{R(r_0)} \, ,
\eeq
thus giving, 
\beq
F(U) \equiv 1 - R(r_0) (r - r_0) =  \l(\frac{R(r_0) U}{2}\r)^2 \, .
\eeq
Substituting the above expression in (\ref{eq:act-lem}), we get
\br 
S_0[{\rm emission}]& =& - \frac{4 E}{R(r_0)}\int_{1 - \e}^{1 + \e}
dU \frac{U}{1 - U} + {\rm real~part} \nonumber \\ 
&=& \frac{4 i \pi E}{R(r_0)} + {\rm real~part} \, ,
\label{eq:emi-lem}
\er
where $U$ is rescaled as $ [R(r_0) U/2] \to U$.

In order to obtain the action for absorption of particles
corresponding to the ingoing particles [$(\pa S_0/\pa r) < 0$], we
have to repeat the above calculation for the metric (\ref{eq:gen-lem})
with ``$+$'' sign of expression (\ref{eq:rel-R-tP}). In this case, it
is easy to see that the only singular solution corresponds to in-going
particles and so
\beq
S_0[{\rm absorption}] = - \frac{4 i \pi E}{R(r_0)} + {\rm real~part} \, .
\label{eq:abs-lem}
\eeq
Constructing the semi-classical propagator in the usual manner and
taking the modulus square we obtain the probability. Now, extending
the double mapping of the paths to the generalized \lem~coordinates
and squaring the modulus, we get the temperature associated with the
quantum scalar fields propagating in the generalized \lem~coordinates
is given by
\beq
\beta^{-1} = \frac{R(r_0)}{4\pi} =  \frac{\kappa(r_0)}{2 \pi}\, ,  
\eeq 
which is same as the expression obtained in generalized
\pl~coordinate. For dS, we get $\beta^{-1} = 1/(2 \pi l)$.

\section{}
\label{app:bogo}

In this appendix, we recover Bogoliubov coefficients (for space-times
with single horizon) solely by the mode functions of the scalar field
near the horizon \cite{srini-pri}.  Let us consider a quantum field
propagating in a spherically symmetric coordinate
(\ref{eq:gen-spher}). The equation of motion of the scalar field $\Phi
= \Psi(t, r) Y_{lm}(\theta,\phi)$ is
\beq
\frac{r^2}{g(r)}\frac{\pa^2 \Psi}{\pa t^2} - \frac{\pa}{\pa
r}\l(r^2 g(r) \frac{\pa \Psi}{\pa r}\r) + L^2 \Psi = 0 \, .
\label{eq:scalar-spher}
\eeq
Since, our interest is near the horizon, we can transform the above
expression in terms of a new variable $x \equiv r - r_0$. The solution
to the scalar field, close to the horizon $x = 0$, is given by
\beq 
\Psi = C_1 \exp(-iE t)~ x^{i E/R(0)},
\label{eq:sol}
\eeq
where $C_1$ is an arbitrary constant. Note that $\Psi^*$ is also a
solution to the scalar field equation.

In the usual field theory description, Bogoliubov transformations 
relate two distinct orthonormal sets of modes which are the solutions 
to the equation of motion. In the case of method of complex paths, 
the two sets of modes correspond to $\Psi(x<0)$ and $\Psi(x>0)$. We can 
write the two sets of modes as
\br
\Psi(x < 0)& = & C_1 \psi(-x) + C_2 \psi^*(-x)\, , \nonumber \\
\Psi(x > 0)& = & C_{\alpha} \psi (x) + C_{\beta} \psi^*(x) \, , 
\label{eq:bogol-main}
\er
where $C_1$, $C_2$, $C_{\alpha}$ and $C_{\beta}$ are constants to be
determined. In the region $x < 0$, $\psi(-x)$ represents an outgoing
particle from the left of the horizon. Implementing the method of
complex paths by rotating in the {\it upper} complex plane, we get
\beq
C_{\alpha} = C_1 \exp[\pi E/R(0)]\, .
\label{eq:bogol1}
\eeq
In the region $x < 0$, $\psi^*(-x)$ represents an ingoing particle to
the left of the horizon. Here again, implementing the method of
complex paths by rotating in the {\it lower} complex plane, we 
obtain
\beq
C_{\beta} = C_2 \exp[-\pi E/R(0)]\, .
\label{eq:bogol2}
\eeq
This leaves us with identifying a relation between $C_1$ and $C_2$.
In order to do this, let us consider the physical situation of
particle production near the black-hole horizon. In this case, a
virtual particle-antiparticle pair to the left of the horizon gets
converted to a real particle-antiparticle pair to the right of the
horizon by the tidal action of the gravitational field implying that
the total current to the left of the horizon is zero. Thus, we get
\beq
|C_1|^2 = |C_2|^2.
\label{eq:bogolcond}
\eeq

This implies,                
\beq
|{C_{\beta}|^2 = |C_{\alpha}}|^2 \exp[-2 \pi E/\kappa].
\label{eq:bogol3}
\eeq
Interpreting $|C_{\alpha}|^2 \equiv |\alpha_E|^2$ and $|C_{\beta}|^2 
\equiv |
\beta_E|^2$ as the Bogoliubov coefficients and using the unitarity
condition, we get
\br
\l|\alpha_E\r|^2 & = & {1 \over 1 -\exp[-2\pi E/\kappa] } \nonumber \\
\l|\beta_E\r|^2  & = & {1 \over \exp[2\pi E/\kappa] - 1}.
\label{eq:bogol4}
\er
These are the well known relations for the Bogoliubov
coefficients.

\end{document}